\documentclass[twocolumn,aps,prl]{revtex4-2}
\usepackage{times}
\usepackage{amsmath,amsfonts,amssymb}
\usepackage{amsthm,mathrsfs}
\usepackage{soul,bm,array,graphicx,bbold,multirow}
\usepackage[normalem]{ulem}
\usepackage[makeroom]{cancel}
\usepackage[usenames,dvipsnames]{xcolor}
\usepackage[colorlinks=true,citecolor=blue,linkcolor=red]{hyperref}
\usepackage{pdfpages}
\usepackage{booktabs} 
\usepackage{wasysym, utfsym}
\usepackage{pifont}
\usepackage{threeparttable}
\usepackage{makecell}
\makeatletter
\AtBeginDocument{\let\LS@rot\@undefined}
\makeatother
\newcolumntype{x}[1]{>{\centering\let\newline\\\arraybackslash\hspace{0pt}}p{#1}}

\DeclareMathAlphabet{\mathbbold}{U}{bbold}{m}{n}

\newcounter{subeqn} %

\renewcommand{\Im}{\operatorname{Im}} 
\makeatletter
\@addtoreset{subeqn}{equation}
\makeatother

\renewcommand{\Im}{\operatorname{Im}}

\begin{document}
	
\title{Complex Frequency Detection in a Subsystem }

\author{Juntao Huang$^1$}
\author{Jiangping Hu$^{2,3,4}$}
\email[Corresponding author: ]{jphu@iphy.ac.cn}
\author{Zhesen Yang$^1$}
\email[Corresponding author: ]{yangzs@xmu.edu.cn}

\affiliation{$^1$ Department of Physics, Xiamen University, Xiamen 361005, Fujian Province, China}
\affiliation{$^2$ Beijing National Laboratory for Condensed Matter Physics and Institute of Physics, Chinese Academy of Sciences, Beijing 100190, China}
\affiliation{$^3$ School of Physical Sciences, University of Chinese Academy of Sciences, Beijing 100190, China}
\affiliation{$^4$ New Cornerstone Science Laboratory, Beijing, 100190, China}
	
\date{\today}
	
\begin{abstract}
In this study, we systematically explore the non-Hermitian skin effect (NHSE) and its associated complex-frequency detection in the context of a frequency-dependent non-Hermitian Hamiltonian. 
This Hamiltonian arises from the self-energy correction of the subsystem and can be calculated exactly within our theoretical model, without the need for any approximations.
Additionally, complex frequency detection, which encompasses complex frequency excitation, synthesis, and fingerprint, enables us to detect the physical responses induced by the complex driving frequencies.
Our calculations reveal that both complex frequency excitation and synthesis are incompatible with the non-Hermitian approximation and are unable to characterize the presence or absence of the NHSE.
In contrast, the complex-frequency fingerprint successfully detects the novel responses induced by the NHSE through the introduction of a double-frequency Green’s function.
Our work provides a platform for studying non-Hermitian physics and their novel response in quantum systems rigorously without relying on any approximations.
\end{abstract}
	
\maketitle

\begin{figure}[t]
	\begin{center}
		\includegraphics[width=0.9\linewidth]{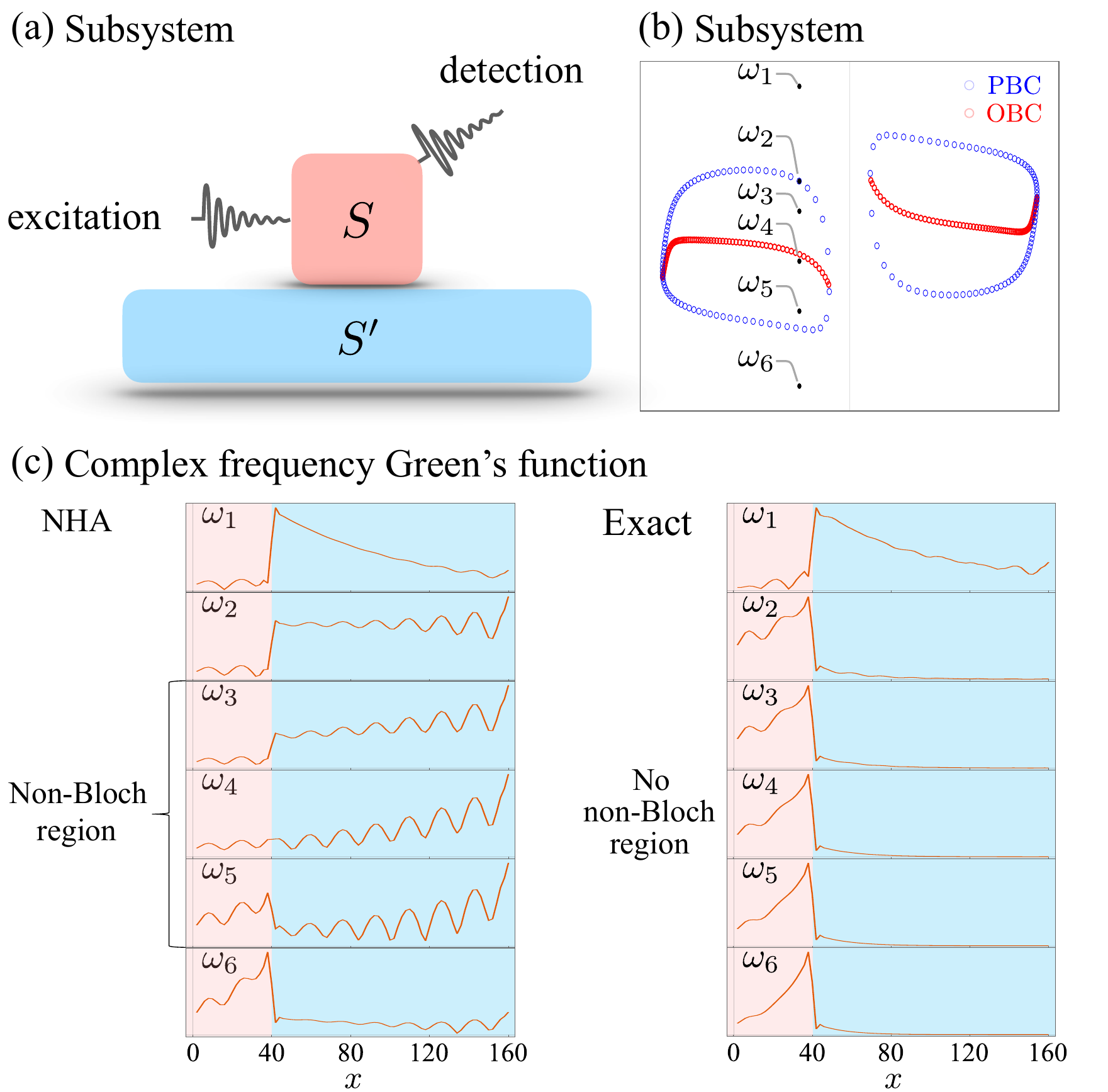}
		\par\end{center}
	\protect\caption{(a) Schematic illustration of the excitation and detection processes within the subsystem $S$, which is coupled to another subsystem $S^{\prime}$. 
	(b) Energy spectra of the approximated non-Hermitian $H_{S,\rm{nH}}$ under PBC and OBC. (c) Comparative analysis of the CFGF obtained with and without the NHA. 
	Detailed model parameters and computational specifications are provided in Fig.\ref{F2}.} 
	\label{F1}
\end{figure}

{\color{red}\em Introduction.}---Non-Hermitian physics, particularly the non-Hermitian skin effect (NHSE), has attracted significant research interest recently~\cite{FoaTorres,Ashida,Rmp93015005,Dingkrev,re-nhse,Lin2023Topo,Prl1210868,PrlYSW,Prl123170401,songfeiprl,hym4,hym5,hym6,hym7,Prb971214,Prl121kunst,PrxUeda,prbedva,lchTR,transferkunst,PrbDY,PrlYM,Prbjh,Prr1023013,Prxkawabata,prlOkuma,Prlborgnia,PrbKOM,prbhigherorder,Prl125126402,PrlYZFH,Prl125186,Prb102085151,prlgjb,Prbzxz,llhcritical,gdse,cw2,cw3,djj}. 
Many experimental developments have been made in classical wave platforms, including acoustic~\cite{acous1,transient,Prb106134112,zxj,lzy,Prl133126601,Prb110L140305,zqc,szq-gh,zhong2024,wang2024disorder,hu2024,yzj}, optical~\cite{scienceaaz8727,xiao-wang,2024NatPh,xpanderson,Prl133073803,Prl133070801,Prl132113802,Prl132063804,Prl130263801,Prl129113601,Prb110094308,Lpx1,vicencio2024}, mechanical~\cite{ghatak,branden,cyy,mgc1,mgc2,mgc3,sciadvadf7299}, and electrical circuit systems~\cite{helbig,Prr20232,Liu-Zhang,zoudy,Prr4033109,shang-tie,Prb107085426,osdbe,higherank,Prr5043034,Prb108035410,Prb107184108,jin2024}.
For quantum systems, a natural theoretical framework involves starting with the standard Green's function (GF) of subsystems~\cite{Prl125186,Koziiprb,Prldmft,PrbKaneshiro,cw1,Prlshaokai}. 
As schematized in Fig.~\ref{F1} (a), when both excitations and responses are confined within a subsystem $S$, a frequency-dependent self-energy emerges, i.e., $\Sigma_S(k,\omega)$. 
Under zeroth-order approximation, i.e., $\Sigma_{S}(k,\omega)\simeq \Sigma_{S,0}$, this yields an effective non-Hermitian Hamiltonian, i.e., $H_{S,{\rm nH}}(k)=H_S(k)+\Sigma_{S,0}$, referred to as the {\em non-Hermitian approximation} (NHA) in this work~\cite{Zhysci,Prlsht,Prldmft,Koziiprb,yoshidaprb2,Rausch_2021,Prbpeter,Bergholtzprr,sanjosesci,philipprb}.

When $H_{S,\rm{nH}}(k)$ exhibits non-zero spectral winding as shown in Fig.~\ref{F1} (b), the corresponding open boundary condition (OBC) system manifests the  NHSE~\cite{prlOkuma,Prl125126402}. 
In order to understand this phenomenon, the generalized Brillouin zone or non-Bloch band theories are developed~\cite{Prl1210868,PrlYM,PrlYZFH,Prl125126402}.
Physical responses governed by Brillouin/generalized Brillouin zones are categorized as Bloch/non-Bloch responses, respectively~\cite{Prl123170401,CFF,GFwz1,GFwz2}. 
For instance, the single-particle GF under OBC, i.e., $G_{\rm OBC}^R(\omega)$, belongs to the non-Bloch responses when $\omega$ possesses non-zero spectral winding and Bloch responses otherwise~\cite{SM}. 
Fig.~\ref{F1} (c) illustrates this statement: frequencies from $\omega_3$ to $\omega_5$ (with non-zero spectral winding) exhibit rightward-dominating divergence in the GF upon excitation at $x=40$.
Such a divergent behavior in the GF is a hallmark of the non-Bloch response and can be used to characterize the presence of point gap topology as well as the NHSE~\cite{prlOkuma,Prl125126402,SM}.

\begin{figure*}[t]
	\begin{center}
		\includegraphics[width=0.95\textwidth]{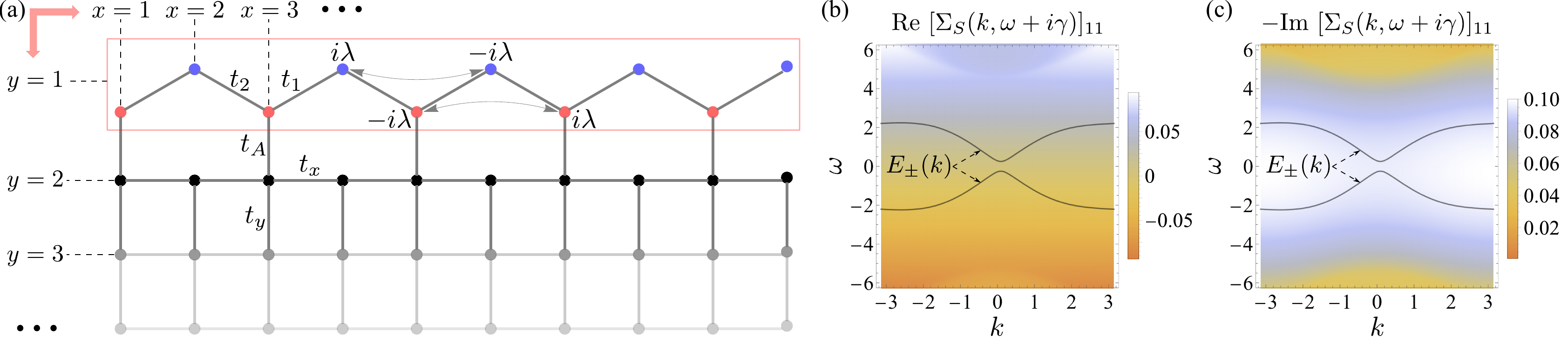}
		\par\end{center}
	\protect\caption{(a) Schematic representation of the tight-binding model for the composite system, with subsystem $S$ explicitly demarcated by the red rectangular enclosure.
	(b) and (c) A comparison between the self-energy and the bare bands of the subsystem $S$, which ensures the validity of the NHA.  
	The parameters include $t_{1}=1.2,t_{2}=-1,\lambda=-1,\mu_{S}=0.2,t_{x}=1,t_{y}=3.5,\mu_{S^{\prime}}=0.3,t_{A}=0.6,\gamma=0.1$, $N_{S}=80$ and $N_{y}=300$.  } 
	\label{F2}
\end{figure*}

In this work, we address the following crucial question: Does the non-Bloch response persist in the exact GF beyond the NHA framework? 
Through a concrete model (detailed later),  Fig.~\ref{F1} (c) (right panel) demonstrates that the unidirectional divergence becomes absent in the exact GF without applying any approximations. 
This result raises three fundamental issues: (i) How to interpret this result? (ii) Does our subsystem truly exhibit the NHSE and non-Bloch responses? (iii) If so, how do we detect them?
Notably, since the frequency $\omega$ in the GF has become complex, it is essential to explore complex frequency detection methods to answer these questions.

Complex frequency detection techniques, comprising the complex frequency excitation (CFE)~\cite{zs1, zs2, zs3, Prl124, Prl109, Prl112, tetikol, Tsakmak, Prl1292022, PRX13}, synthesis (CFS)~\cite{zs1,zs2,CTchanNC}, and fingerprinting (CFF)~\cite{CFF}, offer experimentally viable strategies to extract critical information from the complex-frequency GF (CFGF)~\cite{SM}. 
However, existing studies typically assume a frequency-independent non-Hermitian Hamiltonian, i.e., $H_{S,\rm{nH}}(k)$. 
How to generalize the discussions to the frequency dependent cases, like $H_{S,{\rm eff}}(k,\omega)=H_S(k)+\Sigma_{S}(k,\omega)$? 
As far as we know, this problem has not yet been thoroughly explored in the literature. 
This leads to the second motivation of this work, that is, providing a comprehensive discussion on the complex frequency detections of the exact sub-system effective Hamiltonian, i.e., $H_{S,{\rm eff}}(k,\omega)$. 

Starting from a exactly solvable model, our calculations reveal the following conclusion: while CFE, CFS, and CFF yield consistent outcomes under $H_{S,{\rm nH}}(k)$, they diverge when applied to $H_{S,{\rm eff}}(k,\omega)$. 
Specifically, as shown in Fig.~\ref{F1} (c), the CFE and CFS yield the exact results shown in the right panels (with no non-Bloch response), whereas the CFF detects the approximated ones shown in the left panels (with the presence of non-Bloch response). 
Our work not only establishes a foundation for understanding the NHSE and non-Bloch dynamics in quantum systems but also provides critical guidance for experimental verification.
 
{\color{red}\em Model and subsystem Green's function}---Our Hamiltonian adopts the following Hermitian form: 
\begin{equation}
\hat{H}_{tot}=\hat{H}_{S}+\hat{H}_{S^{\prime}}+\hat{H}_{SS^{\prime}},
\end{equation}
where the hopping parameters are illustrated in Fig.~\ref{F2} (a). 
The subsystem $S$ corresponds to a one-dimensional spinless Rice-Mele model with broken time-reversal symmetry~\cite{Prl125186}, positioned in the $y=1$ region, with sublattices $A$ and $B$ represented by red and blue points, respectively. 
Applying Fourier transformation, we obtain the Bloch Hamiltonian: 
\begin{equation}
	H_{S}(k)=(t_{1}+t_{2}\mathrm{cos}~k)\sigma_{x}+t_{2}\mathrm{sin}~k\sigma_{y}+(\lambda \mathrm{sin}~k-\mu_{S})\sigma_{z},
\end{equation}
where the Pauli matrix acts on the sublattice degree of freedom.

The subsystem $S^{\prime}$ comprises a two-dimensional single-band model inhabiting the $y\geq 2$ region, featuring anisotropic hoppings with $t_{x}\neq t_{y}$. 
 Its $x$-direction Bloch Hamiltonian adopts the layer-resolved form:
\begin{equation}
	H_{S^{\prime}}(k)=\begin{pmatrix}
		h_{S^{\prime}}(k)&t_{y}\sigma_0&0&0&0&...\\
		t_{y}\sigma_0&h_{S^{\prime}}(k)&t_{y}\sigma_0&0&0&...\\
		0&t_{y}\sigma_0&h_{S^{\prime}}(k)&t_{y}\sigma_0&0&...\\
		...&...&...&...&...&...
	\end{pmatrix},
\end{equation}
where $h_{S^{\prime}}(k)=(t_{x}+t_{x}\mathrm{cos}~k)\sigma_{x}+t_{x}\mathrm{sin}~k \sigma_{y}-\mu_{S^{\prime}}\sigma_{0}$ describes the Hamiltonian at each $y$-layer using a two-site unit cell. 
The interlayer hopping in the subsystem $S'$ is parameterized by $t_{y}\sigma_0$, and the total number of $y$-layers in the subsystem $S'$ is denoted as $N_y$. 

For conceptual clarity, we restrict inter-subsystem coupling to the $A$-sublattice (denoted as $t_{A}$). 
In this case, as derived in the Supplementary Material (SM), the exact self-energy for the subsystem $S$ becomes:
\begin{equation}
	\Sigma_S(k,\omega)=\begin{pmatrix}
			\Sigma_{S}^{AA}(k,\omega)&0\\
		0&0\\
	\end{pmatrix},
\end{equation}
where $\Sigma_{S}^{AA}(k,\omega)=t^{2}_{A}[1/(\omega-H_{S^{\prime}}(k))]_{11}$. 
Fig.~\ref{F2} (b) and (c) present the corresponding density plots of the real and imaginary components of $\Sigma^{AA}_{S}(k,\omega)$, respectively.

\begin{figure*}[t]
	\begin{center}
		\includegraphics[width=0.95\textwidth]{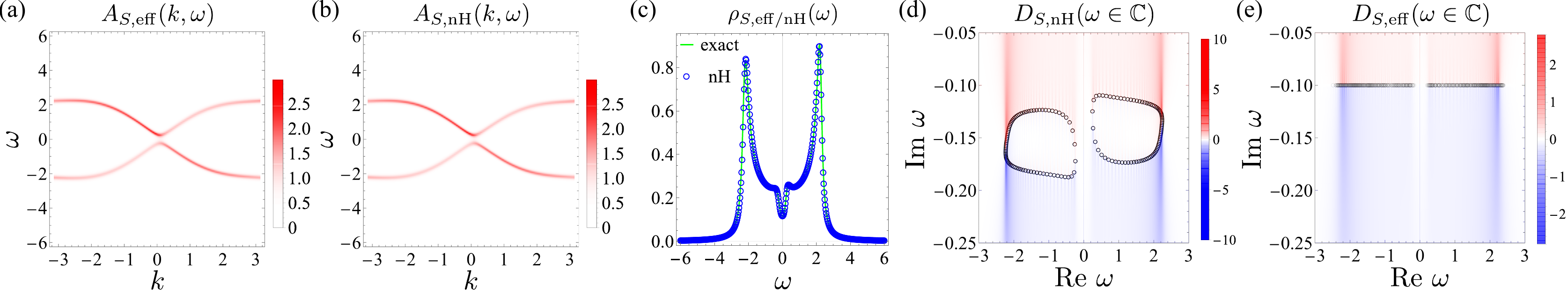}
		\par\end{center}
	\protect\caption{(a) and (b) The spectral functions with and without NHA, respectively. 
	(c) A comparison between the real frequency DOS with and without the NHA.
	(d) and (e) The complex frequency DOS with and without NHA, respectively, where the black circles indicate the poles of the CFGF. 
	The parameters are identical to that in Fig.~\ref{F2}. } 
	\label{F3}
\end{figure*}

Based on this exact self-energy, the exact GF for subsystem $S$ under PBC can be written as:
\begin{equation}
	G_{S,{\rm eff}}^R(k,\omega)=\frac{1}{\omega-H_{S,\rm eff}(k,\omega)},
	\label{E5}
\end{equation}
where the frequency-dependent non-Hermitian Hamiltonian is explicitly expressed as
\begin{equation}
	H_{S,\rm eff}(k,\omega)=H_{S}(k)+\Sigma_S(k,\omega+i\gamma)-i\gamma.
\end{equation}
Here, the parameter $i\gamma$ follows the conventional retarded GF formalism, while $\omega$ spans the entire complex energy plane. 
In practical numerical calculations, $\gamma$ is chosen as a small value, which does not affect the qualitative results.

{\color{red}\em Non-Hermitian approximation}---We now check the validity of the NHA.
As shown in Fig.~\ref{F2} (b) and (c), the energy band of $H_S(k)$ is plotted against the real and imaginary components of the self-energy, respectively. 
Notably, within the band dispersion region, the self-energy can be well approximated by a constant, i.e., 
\begin{equation}
\Sigma_{S}^{AA}(k,\omega+i\gamma)\approx \Sigma_{S}^{AA}(0,i\gamma)+\mathcal{O}(k)+\mathcal{O}(\omega).
\end{equation}
This NHA yields the following approximated GF, 
\begin{equation}
	G_{S,{\rm nH}}^R(k,\omega)=\frac{1}{\omega-H_{S,\rm nH}(k)},
\end{equation}
where the frequency-independent non-Hermitian Hamiltonian is given by:
\begin{equation}
	H_{S,\rm nH}(k)=H_S(k)+\begin{pmatrix}
		\Sigma_{S}^{AA}(0,i\gamma)&0\\
		0&0\\
	\end{pmatrix}-i\gamma.
\end{equation}

To further verify the NHA, we present a comparative analysis of exact versus approximated spectral functions, i.e., $A_{S,\rm eff/nH}(k,\omega)=-(1/\pi)\mathrm{Im}~\mathrm{Tr}~G^{R}_{S,\rm eff/nH}(k,\omega\in\mathbb{R})$.
As illustrated in Fig.~\ref{F3} (a) and (b), the remarkable agreement between these spectral functions confirms the validity of the NHA. 
This conclusion is further corroborated by the density of states (DOS) at real frequencies shown in Fig.~\ref{F3} (c), which follows the conventional definition,  i.e.,  $\rho_{S,\rm{eff}/\rm{nH}}(\omega)=(1/N_{S})\sum_{k}A_{S,\rm{eff}/\rm{nH}}(k,\omega)$.

However, significant discrepancies emerge when considering the complex frequency domain. 
Fig.~\ref{F1} (c) demonstrates a representative example of these differences. 
For comprehensive analysis, we introduce the complex-frequency DOS~\cite{CFF} across the entire complex plane, i.e., 
\begin{equation}
	D_{S,\rm eff/nH}(\omega)=-\frac{1}{N_{S}\pi}\sum\limits_{k}\mathrm{Im}~\mathrm{Tr}~G^{R}_{S,\rm eff/nH}(k,\omega\in\mathbb{C}),
\end{equation}
which can reveal the pole structures of the CFGF~\cite{CFF,SM}. 
As shown in Fig.~\ref{F3} (d)-(e), the approximated GF displays two closed-loop poles, contrasting with the exact case that features a branch cut positioned at $\mathrm{Im}~\omega=-\gamma$. 
Due to the vanishing of the spectral winding in the exact poles under PBC, it is expected that there is no non-Bloch response in the GF, which explains the paradox observed in Fig.~\ref{F1} (c).

In summary, although the NHA demonstrates excellent performance in simulating real-frequency GF, it fundamentally modifies the global pole structures. 
Therefore, it is necessary to examine the nontrivial roles of the NHA in variant complex-frequency detection methodologies.

\begin{figure*}[t]
	\begin{center}
		\includegraphics[width=0.95\textwidth]{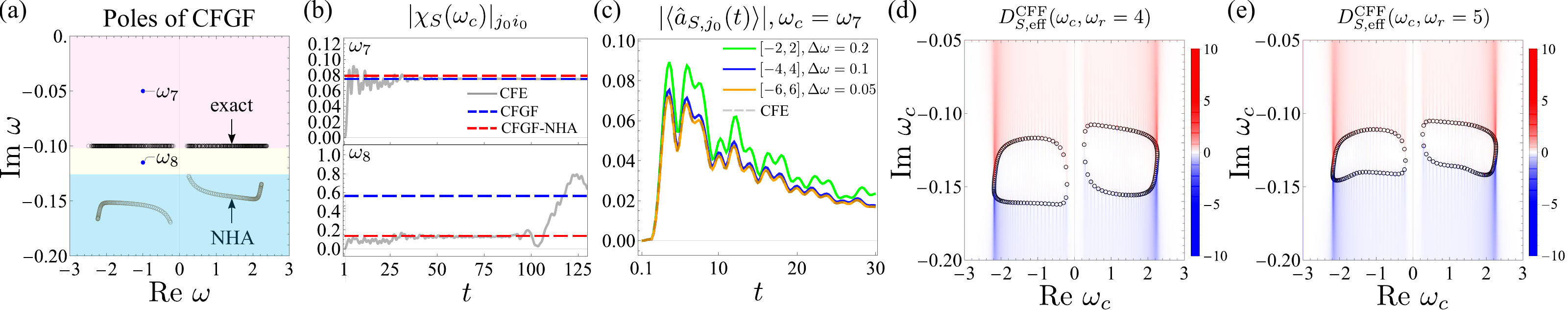}
		\par\end{center}
	\protect\caption{(a) The poles of the OBC subsystem GF with and without NHA. 
	(b) A comparison between the time-dependent response function for the CFE and the CFGF with and without NHA. 
	Here $\omega_{7}=-1-0.05i$, $\omega_{8}=-1-0.115i$, $j_{0}=30$, and $i_{0}=36$.
	(c) A comparison between the CFE and CFS at $\omega_{7}$ with different synthesis domains and spectral resolution. Here $i_{0}=36$.
	(d) and (e) The complex frequency DOS for the CFF at $\omega_{r}=4$ and $\omega_{r}=5$, respectively. 
	Other parameters are identical to that in Fig.~\ref{F2}. } 
	\label{F4}
\end{figure*}

{\color{red}\em Complex frequency detection}---We now focus on the complex frequency detection of the subsystem. 
The theoretical framework originates from the following driving dissipative equation~\cite{CFF}:
\begin{equation}
i\frac{d\langle \hat{\boldsymbol{a}}(t)\rangle}{dt}=(H_{tot}-i\gamma)\langle \hat{\boldsymbol{a}}(t)\rangle+\boldsymbol{F}(t),\label{E8}
\end{equation}
where $H_{tot}$ represents the first quantized Hamiltonian of $\hat{H}_{tot}$ in real space. 
The parameter $\gamma$ characterizes the uniform dissipation within the GF. 
$\langle \hat{\boldsymbol{a}}(t)\rangle=\{\langle \hat{\boldsymbol{a}}_S(t)\rangle,\langle \hat{\boldsymbol{a}}_{S'}(t)\rangle\}^T$ describes the mean value of single-particle operators at each lattice site of the total system, and $\boldsymbol{F}(t)$ denotes the external drive. 
Experimentally, when $\bm{F}(t)$ is applied to the subsystem $S$, the corresponding response $\langle \hat{\boldsymbol{a}}_S(t)\rangle$ is detected.
For example, consider an external driving defined as
\begin{equation}
	\bm{F}(t)=\theta(t)e^{-i\omega_r t}\{0,..,0,F_{S,i},0,...,0\}^{T}, ~~~~\omega_r\in\mathbb{R},
	\label{E9}
\end{equation}
which acts exclusively on the $i$th-site of the subsystem $S$. 
Following SM derivations, the induced response at the $j$-th site, i.e., $\langle \hat{a}_{S,j}(t)\rangle$, demonstrates proportionality to the real-frequency Green's function in the long-time limit. 
This relationship can be expressed as:
\begin{equation}
	\lim_{t\rightarrow\infty}\langle \hat{a}_{S,j}(t)\rangle=\left[G_{S,{\rm eff}}^R(\omega_r)\right]_{ji}F_{S,i}e^{-i\omega_rt},
\end{equation}
where $G_{S,\rm eff}^R(\omega_r)$ denotes the exact GF of the subsystem defined in Eq.~\ref{E5} (real space representation), and $\omega_r$ corresponds to the driving frequency. 
Given the predetermined values of $F_{S,i}$ and $\omega_r$, the real-frequency Green's function becomes experimentally accessible through $\langle \hat{a}_{S,j}(t)\rangle$ detection. 
This methodology maintains alignment with classical wave system protocols for real-frequency Green's function measurement~\cite{Haus,Huang94,Li10,Fan03,Fan04}. 

{\color{red}\em Complex frequency excitation}---When extending the driving frequency from real ($\omega_r$) to complex values ($\omega_c$) in Eq.~\ref{E9}, the concept of CFE emerges. 
As demonstrated in the SM, the solution to Eq.~\ref{E8} under complex frequency driving
\begin{equation}
	\bm{F}(t)=\theta(t)e^{-i\omega_c t}\{0,..,0,F_{S,i},0,...,0\}^{T}, ~~~~\omega_c\in\mathbb{C},\label{E14}
\end{equation}
takes the form
\begin{equation}
	\langle \hat{a}_{S,j}(t)\rangle=\left[\chi_{S}(\omega_c)\right]_{ji}F_{S,i}e^{-i\omega_ct},
\end{equation}
where the response function  $\chi_{S}(\omega_c)$ is expressed as
\begin{equation}\begin{aligned}
	\chi_{S}(\omega_c)=&G^{R}_{S,\rm eff}(\omega_{c})-[G^{R}_{tot}(\omega_{c})e^{-i(H_{tot}-i\gamma-\omega_{c})t}]_{S}.
\label{E13}
\end{aligned}\end{equation}
Here, the subscript $S$ indicates the subsystem component, and $G^{R}_{tot}(\omega_{c})$ denotes the CFGF for the total system. 
We note that this solution assumes the initial condition $\langle \hat{\boldsymbol{a}}(0)\rangle=0$, whose value will not affect the subsequent conclusions, as proved in the SM. 

A critical observation arises from the above formulation: when the complex frequency $\omega_{c}$ resides above the branch cut determined by $i\gamma$ (light pink region in Fig.~\ref{F4} (a)), the corresponding response function will converge to the exact CFGF of the subsystem in the $t\rightarrow \infty$ limit, i.e.,
\begin{equation}
	\lim_{t\rightarrow\infty}\chi_{S}(\omega_c)=G_{S,\rm eff}^R(\omega_c),~~~~\Im \omega_c>-\gamma.
\end{equation}
Conversely, frequencies below the $i\gamma$ line will yield divergent responses due to the second term's divergence. 
As numerically verified in Fig.~\ref{F4} (b), only the $\omega_7$ case converges to the exact CFGF (blue dashed lines). 

For comparative analysis, we show the NHA-predicted CFGF as red dashed lines, which exhibit clear deviations from our calculated response functions. This discrepancy stems from the modification of GF pole structures under the NHA (Fig.~\ref{F3} (d)-(e)), rendering it invalid for frequencies distant from the real axis. 
Fig.~\ref{F4} (b) explicitly demonstrates this limitation: for $\omega_7$, converged results deviate from NHA predictions; $\omega_8$ exhibits response divergence despite NHA suggesting convergence.

Key conclusions are summarized in Table~\ref{T1}:
(i) The CFE exclusively detects the exact CFGF, i.e., Eq.~\ref{E5}, above the total Hamiltonian's imaginary gap (light pink region);
(ii) NHA fails in regions far from the real axis;
(iii) Absence of non-Bloch response in long-time dynamics due to vanishing of spectral winding number. 

\begin{table}[b]
\begin{threeparttable}
\centering
\caption{Results of CFE, CFS, and CFF for $t\rightarrow \infty$}
\setlength{\arrayrulewidth}{1pt}
\renewcommand{\arraystretch}{1.3} 
\setlength{\tabcolsep}{9pt}
\begin{tabular}{c|c|c|c|c}
\hline 
 & \makecell{Effective\\ Region}& CFGF &NHA &\makecell{Non-Bloch\\ Resonse}\\
\hline
CFE & $\mathrm{Im}~\omega_{c}>\Gamma\tnote{*} $&$\rm{Eq.}~\ref{E5}$&\raisebox{0.2ex}{\scalebox{0.85}[1]{$\times$}}&\raisebox{0.2ex}{\scalebox{0.85}[1]{$\times$}}\\
\hline
CFS & \multicolumn{4}{c}{the same as CFE}\\
\hline
CFF&$\omega_{r}\in\mathbb{R}$&$\rm{Eq.}~\ref{E20}$&\raisebox{0.6ex}{\scalebox{0.7}{$\sqrt{}$}}&\raisebox{0.6ex}{\scalebox{0.7}{$\sqrt{}$}}\\
\hline
\end{tabular}
\label{T1}
\begin{tablenotes}
\item[*]{$\Gamma$ characterizes the imaginary gap. In the exact case, $\Gamma=-\gamma$.}
\end{tablenotes}
\end{threeparttable}
\end{table}

{\color{red}\em Complex frequency synthesis}---We now examine the CFS, a method based on the following observation: within the framework of CFE, the nonzero component of $\bm{F}(t)$ in Eq.~\ref{E14} can be transformed through Fourier analysis as
\begin{equation}
	\theta(t)e^{-i\omega_c t}=\frac{1}{2\pi}\int_{-\infty}^{\infty}\frac{i}{\omega-\omega_c}e^{-i\omega t}d\omega,
\end{equation}
where $\mathrm{Im}~\omega_{c}<0$ and $t>0$.
Experimentally measured real frequency responses enable comprehensive reconstruction of the CFE solution through signal synthesis:
\begin{equation}
	\langle \hat{\boldsymbol{a}}_{S,\omega_{c}}(t)\rangle=\frac{i}{2\pi}\sum\limits^{\omega_{b}}_{\omega=\omega_{a}}\frac{\Delta \omega}{\omega-\omega_{c}}\langle \hat{\boldsymbol{a}}_{S,\omega}(t)\rangle,  
	\label{E18}
\end{equation}
where $[\omega_{a},\omega_{b}]$ defines the synthesis domain and $\Delta \omega$  specifies the spectral resolution.

As summarized in Tab.~\ref{T1}, CFS demonstrates theoretical equivalence with CFE. 
A quantitative validation is presented in Fig.~\ref{F4} (c), where the result derived from CFE is compared with synthesized responses across progressively expanding synthesis domains $\Delta\omega$ and refined spectral resolutions. 
The convergence toward the CFE result confirms the validity of the CFS method. 

{\color{red}\em Complex frequency fingerprint}---To capture the non-Bloch response, the recently developed CFF method under real-frequency driving is necessary. 
As demonstrated in the SM, as $t\rightarrow \infty$, the CFF specifically measures the following double-frequency GF,
\begin{equation}
	G^{\rm CFF}_{S,\rm eff}(k,\omega_c,\omega_r)=\frac{1}{\omega_c-H_{S,\rm eff}(k,\omega_r)},\label{E20}
\end{equation}
where $\omega_c\in\mathbb{C}$ represents a complex parameter and $\omega_r\in\mathbb{R}$ corresponds to the driving frequency defined in Eq.~\ref{E9}. 
Crucially, for any fixed driving parameter $\omega_r$, this double-frequency GF precisely corresponds to the CFGF associated with the non-Hermitian Hamiltonian $H_{S,\rm eff}(k,\omega_r)$.
As $\omega_r$ evolves, the corresponding double-frequency GF changes accordingly. 
Fig.~\ref{F4} (d) and (e) exemplify this behavior through the following complex-frequency DOS
\begin{equation}
	D_{S,\rm eff}^{\mathrm{CFF}}(\omega_{c},\omega_{r})=-\frac{1}{N_{S}\pi}\sum_{k}\mathrm{Im}~\mathrm{Tr}~G^{\mathrm{CFF}}_{S,\rm{eff}}(k,\omega_{c},\omega_{r})
\end{equation}
at frequencies $\omega_r=4$ and $\omega_{r}=5$, respectively.

For any given $\omega_r$, once $H_{S,\rm eff}(k,\omega_r)$ exhibits the NHSE, the corresponding non-Bloch response can be detected via the double frequency GF by setting $\omega_c$ with a nonzero spectral winding. 
An example is illustrated on the left side of Fig.~\ref{F1} (c) with $\omega_r=0$. 

In summary, as illustrated in Tab.~\ref{T1}:
(i) The CFF methodology effectively identifies double-frequency GF governed by Eq.~\ref{E20};
(ii) For each driving frequency $\omega_r$, the CFF is described by the non-Hermitian Hamiltonian at $\omega_r$, i.e., $H_{S,\rm eff}(k,\omega_r)$; 
(iii) This framework establishes a unified platform for the systematic exploration of NHSE and non-Bloch responses in a Hermitian quantum system. 

{\color{red}\em Conclusion}.---In this work, we have established a framework for detecting complex frequency responses in a frequency-dependent Hamiltonian of the subsystem. 
Our comparative study reveals that while both CFE and CFS methods yield identical results characterizing the exact CFGF without NHSE signatures, the CFF approach uniquely manifests non-Bloch responses through its detection of double-frequency GFs. 
Our work elucidates the physical origin of non-Hermitian Hamiltonians in quantum systems through exact theoretical formalisms, bypassing traditional approximation constraints. 
These advancements provide foundational insights for decoding non-Hermitian phenomena in quantum systems and offer concrete guidance for experimental implementations.

J. Huang and Z. Yang were sponsored by the National Key R$\&$D Program of China (No. 2023YFA1407500) and the National Natural Science Foundation of China (No. 12322405, 12104450, 12047503). 
J. Hu was sponsored by the Ministry of Science and Technology (Grant No. 2022YFA1403901), National Natural Science Foundation of China ( No. 12494594), and the New Cornerstone Investigator Program.

	\bibliography{ref}
	\bibliographystyle{apsrev4-1}
	
\end{document}


\title{Supplementary Information for ``Complex Frequency Detection in a Subsystem"}

\author{Juntao Huang$^1$}
\author{Jiangping Hu$^{2,3,4}$}
\author{Zhesen Yang$^1$}

\affiliation{$^1$ Department of Physics, Xiamen University, Xiamen 361005, Fujian Province, China}
\affiliation{$^2$ Beijing National Laboratory for Condensed Matter Physics and Institute of Physics, Chinese Academy of Sciences, Beijing 100190, China}
\affiliation{$^3$ School of Physical Sciences, University of Chinese Academy of Sciences, Beijing 100190, China}
\affiliation{$^4$ New Cornerstone Science Laboratory, Beijing, 100190, China}


\maketitle


\tableofcontents

\newpage
\section{Divergent behavior of the Green's function}
This section demonstrates that the divergence of the complex frequency Green's function is uniquely associated with the existence of NHSE.
Since a non-zero winding number characterizes the NHSE, this section concisely verifies that a non-zero winding number leads to the non-Bloch response or the divergent behavior of the Green's function, consistent with the conclusions presented in Fig. 1.

To examine the Green's function, we introduce $\beta=e^{ik}$ and label $\beta_{n=1,...,2M}$ as the roots of $\mathrm{det}[\omega-H(\beta)]=\frac{P(\beta,\omega)}{\beta^{M}}=0$, where $M$ represents the orbital degrees of freedom. For a given complex frequency $\omega$, the roots can be ordered as $|\beta_{1}(\omega)|\leq |\beta_{2}(\omega)|\leq|\beta_{M}(\omega)|\leq |\beta_{M+1}(\omega)|\leq...\leq |\beta_{2M}(\omega)|$, where $\beta_{1},\beta_{2},...\beta_{M}$ are enclosed by the generalized Brillouin Zone~\cite{GFwz1,GFwz2}. The scaling behavior of the Green's function under OBC is described as~\cite{GFwz1,GFwz2}
\begin{equation}
	[G_{\mathrm{OBC}}]_{i_{\alpha},j_{\gamma}}\sim \begin{cases}
		\beta^{-(j-i)}_{M+1}, & i<j,\\
		\beta^{i-j}_{M}, & i>j,
	\end{cases}
\end{equation}
for large $|i-j| $, where $\alpha$ and $\gamma$ represent orbital indices.

Additionally, the winding number is given by~\cite{Prl125186}
\begin{equation}
\nu(\omega)=\frac{1}{2\pi i}\int^{2\pi}_{0}dk \partial_{k}\mathrm{ln}~\mathrm{det}[H_{\mathrm{nH}}(k)-\omega],
\end{equation}
where $H_{\mathrm{nH}}(k)$ denotes a non-Hermitian Hamiltonian under PBC (in our case, $H_{\mathrm{nH}}(k)=H_{S,\mathrm{nH}}(k)$).
When $\omega$ lies outside the point gap, the roots satisfy $|\beta_{1}(\omega)|\leq |\beta_{2}(\omega)|<...<|\beta_{M}(\omega)|< 1 < |\beta_{M+1}(\omega)|\leq...\leq |\beta_{2M}(\omega)|$~\cite{GFwz1,GFwz2}. Hence, the inequalities $|\beta_{M}(\omega)|<1$ and $|\beta_{M+1}(\omega)|>1$ hold for all $\nu(\omega)=0$. In this case, we conclude that the scaling behavior of OBC Green's function will exponentially decay as $i$ moves away from $j$, as illustrated in the exact result and NHA results with $\nu(\omega)=0$ in Fig. 1 (c).

When $\omega$ lies inside the point gap, i.e., $\nu(\omega)=1$, it follows that $|\beta_{M}(\omega)|>1$ in our case ($M=2$), indicating that the OBC Green's function exhibits divergent behavior for $i\gg j$, as illustrated in the non-Bloch region in Fig. 1 (c). Thus, the NHSE is uniquely responsible for the divergent behavior of the Green's function.
%

\section{Complex frequency detection methods}
This section reviews how complex frequency techniques, including CFE, CFS, and CFF, can be used to detect the CFGF.

\subsection{The solution of the driving dissipative equation}
The general solution of Eq. 11 can be expressed as
\begin{equation}
\langle \hat{\boldsymbol{a}}(t)\rangle=e^{-i(H_{tot}-i\gamma)t}[-i\int^{t}_{0}e^{i(H_{tot}-i\gamma)\tau}\boldsymbol{F}_{}(\tau)d\tau+\langle \hat{\boldsymbol{a}}(0)\rangle],
\end{equation}
where the external drive $\boldsymbol{F}(t)=\theta(t)e^{-i\omega t}\{F_{1}(0),...,F_{N}(0)\}^{T}$ is considered for general $\omega\in\mathbb{C}$, with $\theta(t)$ representing the step function. A straightforward integral over $\tau$ yields:
\begin{equation}
\begin{split}
\langle \hat{\boldsymbol{a}}_{\omega}(t)\rangle&=e^{-i(H_{tot}-i\gamma)t}[-i\int^{t}_{0}e^{i(H_{tot}-i\gamma)\tau}\boldsymbol{F}(\tau)d\tau+\langle \hat{\boldsymbol{a}}(0)\rangle]\\
&=\frac{e^{-i\omega t}-e^{-i(H_{tot}-i\gamma)t}}{\omega+i\gamma-H_{tot}}\theta(t)\{F_{1}(0),...,F_{N}(0)\}^{T}+e^{-i(H_{tot}-i\gamma)t}\langle\hat{\boldsymbol{a}}(0)\rangle\\
&=G^{R}(\omega)(1-e^{-i(H_{tot}-i\gamma-\omega)t})\boldsymbol{F}(t)+e^{-i(H_{tot}-i\gamma-\omega)t}\mathrm{Diag}\big(\frac{\langle \hat{a}_{1}(0)\rangle}{F_{1}(0)},...,\frac{\langle \hat{a}_{N}(0)\rangle}{F_{N}(0)}\big)\boldsymbol{F}(t),\\
\end{split}
\end{equation} 
where $G^{R}(\omega)=\frac{1}{\omega+i\gamma-H_{tot}}$ and $\mathrm{Diag}(...)$ denotes the diagonal matrix. Notably, for $F_{n}(0)=0$, the divergence ($\sim 1/(F_{n}(0))$) in the Diag(...) term is canceled by the $F_{n}(0)$ in $\boldsymbol{F}(t)$. Subsequently, the response function is determined as:
\begin{equation}
\langle \hat{\boldsymbol{a}}_{\omega}(t)\rangle=\chi(\omega,t)\boldsymbol{F}(t),
\end{equation}	
with
\begin{equation}\begin{split}
	\chi_{mn}(\omega,t)&=G^{R}_{mn}(\omega)-[G^{R}(\omega)e^{-i(H_{tot}-i\gamma-\omega)t}]_{mn}
	+[e^{-i(H_{tot}-i\gamma-\omega)t}]_{mn}\langle \hat{a}_{n}(0)\rangle/F_n(0), \quad F_{n}(0)\neq 0.
		\end{split}
\end{equation}
Here, the response function is well-defined when considering only the components $\chi_{mn}$ with $F_{n}(0)\neq 0$ and random $m$. Moreover, when excitation and detection are performed within subsystem $S$, we obtain 
\begin{equation}
\chi_{S,ji}(\omega,t)=[G^{R}_{S,\mathrm{eff}}(\omega)]_{ji}-[G^{R}(\omega)e^{-i(H_{tot}-i\gamma-\omega)t}]_{ji}
	+[e^{-i(H_{tot}-i\gamma-\omega)t}]_{ji}\langle \hat{a}_{S,i}(0)\rangle/F_{S,i}(0)
\end{equation}
for $i,j\in S$ and $F_{S,i}(0)\neq 0$, where $G^{R}_{S,\mathrm{eff}}(\omega)=\frac{1}{\omega+i\gamma-H_{S}-\Sigma_{S}(\omega+i\gamma)}$, $H_{S}$ denote the matrix form of $\hat{H}_{S}$ in real space under OBC, and $\Sigma_{S}(\omega)$ denotes the self-energy. In this case, the component of the operator mean value takes the form $\langle \hat{a}_{S,j}(t)\rangle=\chi_{S,ji}(\omega,t)\theta(t)F_{S,i}e^{-i\omega t}$ as shown in Eq. 15 of the main text. By adjusting $F_{i}(0)$ or $F_{S,i}(0)$, i.e., the location of excitation, the complete information of $\chi(\omega,t)$ or $\chi_{S}(\omega,t)$ can be obtained. To simplify the following discussion, the initial condition $\langle \hat{\boldsymbol{a}}(0)\rangle=0$ is assumed, which does not affect the qualitative findings.

\subsection{Complex frequency excitation and synthesis}
The CFE method involves the response to an external drive $\boldsymbol{F}(t)$ with temporal attenuation, i.e., $\mathrm{Im}~\omega<0$. The previous study has discussed the case of an entire non-Hermitian system, which corresponds to substituting the total Hamiltonian $H_{tot}-i\gamma$ (with trivial spectral topology under PBC) with a non-Hermitian Hamiltonian $H_{\mathrm{nH}}$~\cite{CFF}. Then, the response function can be expressed by
\begin{equation}
\chi(\omega,t)=G(\omega)(1-e^{-i(H_{\mathrm{nH}}-\omega)t}),\label{S8}
\end{equation}
where $G(\omega)=\frac{1}{\omega-H_{\mathrm{nH}}}$. 
In a dissipative system, if we denote $E_{s}$ as the eigenvalue of $H_{\mathrm{nH}}$ that is the closest to the real axis~\cite{CFF}, then for $\mathrm{Im}~\omega>\mathrm{Im}~E_{s}$, we obtain $\lim\limits_{t\rightarrow \infty}\chi(\omega,t)=G(\omega)$. However, for $\mathrm{Im}~\omega<\mathrm{Im}~E_{s}$, the virtual gain~\cite{zs1, zs2, zs3, Prl124, Prl109, Prl112, tetikol, Tsakmak, Prl1292022, PRX13} arising from the time-dependent contribution covers the information of the Green's function, leading to divergence in the response function. Therefore, $|\mathrm{Im}~E_{s}|$ serves as the imaginary gap of $H_{\mathrm{nH}}$, characterizing the effective region for applying the CFE method. 

For the subsystem $S$, it is evident that the response function is given by
\begin{equation}
\chi_{S,ji}(\omega,t)=[G^{R}_{S,\mathrm{eff}}(\omega)]_{ji}-[G^{R}(\omega)e^{-i(H_{tot}-i\gamma-\omega)t}]_{ji}.
\end{equation}
We find that $\lim\limits_{t\rightarrow \infty}\chi_{S,ji}(\omega,t)=[G^{R}_{S,\mathrm{eff}}(\omega)]_{ji}$ for $\mathrm{Im}~\omega>-\gamma$. However, when $\mathrm{Im}~\omega<-\gamma$, the subsystem response function diverges as $t$ evolves to infinity. Notably, this equation is identical to Eq.~\ref{S8} when the non-Hermitian approximation is applied to subsystem $S$, with $H_{\mathrm{nH}}$ defined as $H_{S}-i\gamma+\Sigma_{S}(\omega_{r}+i\gamma)$, as discussed in the main text.

Besides, in dissipative systems, the detection of the response to a signal with temporal attenuation may pose a challenge in the resolution due to the low signal-to-noise ratio at a long time, hence, the method of CFS is proposed to overcome this issue~\cite{zs1,zs2}. The basic principle of CFS is that a signal of the drive with temporal attenuation can be Fourier transformed into a series of periodic signals with varying real frequencies:
\begin{equation}
\boldsymbol{F}(t)=\frac{1}{2\pi}\int^{\infty}_{-\infty}\boldsymbol{f}(\omega)e^{-i\omega t}d\omega,
\end{equation}
where $t=0$ is considered as the initial state.
Due to the existence of a steady state under periodic signals, the response to a signal with complex frequency can be recovered through a series of responses to real frequency drives without worrying about the resolution problem. 
For example, consider a series of real frequency drives $\boldsymbol{F}_{\omega}(t)=e^{-i\omega t}\{F_{1}(0),F_{2}(0),...,F_{N}(0)\}^{T}$. By choosing $\boldsymbol{f}(\omega)=-i\frac{1}{\omega_{c}-\omega}\{F_{1}(0),F_{2}(0),...,F_{N}(0)\}^{T}$, we obtain
\begin{equation}
\begin{split}
\frac{1}{2\pi}\int^{\infty}_{-\infty}\boldsymbol{f}(\omega)e^{-i\omega t}d\omega&=-\frac{i}{2\pi}\int^{\infty}_{-\infty}\frac{1}{\omega_{c}-\omega}\boldsymbol{F}_{\omega}(0)e^{-i\omega t}d\omega\\
&=\{F_{1}(0),F_{2}(0),...,F_{N}(0)\}e^{-i\omega_{c}t}\\
&=\boldsymbol{F}_{\omega_{c}}(t),
\end{split}
\end{equation}
which represents a complex frequency drive.
For the dissipative non-Hermitian system $H_{\mathrm{nH}}$, the operator mean value under a real-frequency drive can be solved as $\langle \hat{\boldsymbol{a}}_{\omega}(t)\rangle=G(\omega)(1-e^{-i(H_{\mathrm{nH}}-\omega)t})\boldsymbol{F}_{\omega}(t)$. Applying the residue theorem yields
\begin{equation}
\begin{split}
-\frac{i}{2\pi}\int^{\infty}_{-\infty}\frac{1}{\omega_{c}-\omega}\langle \hat{\boldsymbol{a}}_{\omega}(t)\rangle d\omega&=\frac{-i}{2\pi}\int^{\infty}_{-\infty} \frac{1}{\omega_{c}-\omega}G(\omega)(1-e^{-i(H_{\mathrm{nH}}-\omega)t})\boldsymbol{F}_{\omega}(0)e^{-i\omega t}d\omega\\
&=G(\omega_{c})(1-e^{-i(H_{\mathrm{nH}}-\omega_{c})t})\boldsymbol{F}_{\omega_{c}}(t)\\
&=\langle \hat{\boldsymbol{a}}_{\omega_{c}}(t)\rangle.\\
\end{split}
\end{equation}
This result can be written as 
\begin{equation}
\langle \hat{\boldsymbol{a}}_{\omega_{c}}(t)\rangle=\frac{1}{2\pi}\int^{\infty}_{-\infty}f(\omega)\langle \hat{\boldsymbol{a}}_{\omega}(t)\rangle d\omega,
\end{equation}
where $f(\omega)=\frac{i}{\omega-\omega_{c}}$. Importantly, in deriving the above result, the residue theorem is applied only once, and the term $G(\omega)(1-e^{-i(H_{\mathrm{nH}}-\omega)t})$ does not introduce additional poles. Expanding this term, 
\begin{equation}
\begin{split}
G(\omega)(1-e^{-i(H_{\rm{nH}}-\omega)t})&=G(\omega)[i(H_{\rm{nH}}-\omega)t-\frac{1}{2}(-it)^{2}(H_{\rm{nH}}-\omega)^{2}+...]\\
&=-it+\frac{t^{2}}{2}(\omega-H_{\rm{nH}})+...,\\ 
\end{split}
\end{equation}
which confirms that it does not contribute to the contour integral due to the absence of poles when applying the residue theorem.

Therefore, for subsystem $S$, the CFS method can also be applied, leading to the following identity
\begin{equation}
\langle \hat{\boldsymbol{a}}_{S,\omega_{c}}(t)\rangle=\frac{1}{2\pi}\int^{\infty}_{-\infty}f(\omega)\langle \hat{\boldsymbol{a}}_{S,\omega}(t)\rangle d\omega.
\end{equation}
The CFS results discussed above demonstrate their equivalence with the CFE method. However, both methods cannot detect the non-Bloch response in subsystems without non-Hermitian approximation.

\begin{figure*}[b]
	\begin{center}
		\includegraphics[width=0.9\textwidth]{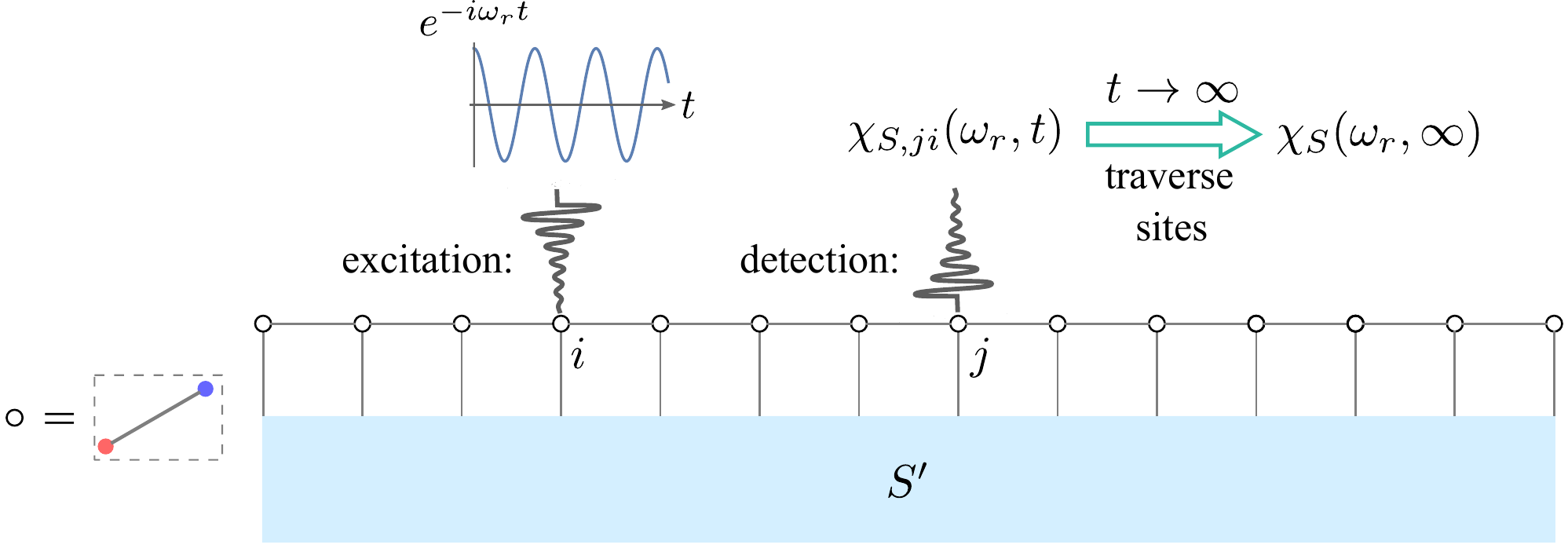}
		\par\end{center}
	\protect\caption{The schematic diagram of the CFF method applied to subsystem $S$. Here, the traversed sites $i,j$ include the orbital degrees, as illustrated by the red and blue circles in our model.} 
	\label{S1}
\end{figure*}

\subsection{Complex frequency fingerprint}
According to the response function to a real frequency drive, the CFF for the entire system is defined as~\cite{CFF}
\begin{equation}
\mathscr{G}_{\omega_r}(t)=\frac{1}{\omega_{c}-\omega_r+\chi^{-1}(\omega_{r},t)}.
\end{equation}
In the non-Hermitian system $H_{\mathrm{nH}}$, since $\lim\limits_{t\rightarrow \infty}\chi^{-1}(\omega_{r},t)=\omega_{r}-H_{\mathrm{nH}}$, we obtain $\lim\limits_{t\rightarrow \infty}\mathscr{G}_{\omega_r}(t)=1/(\omega_{c}-H_{\mathrm{nH}})$, independent of the choice of $\omega_r$. Consequently, by tuning the complex parameter $\omega$ across the entire complex plane, the complex frequency Green's function can be obtained. In this case, the non-Hermitian Hamiltonian is frequency-independent, which is consistent with the observable detected by the CFE and CFS methods.

The CFF for the subsystem can analogously be defined using the subsystem response function:
\begin{equation}
\mathscr{G}_{S,\omega_r}(t)=\frac{1}{\omega_{c}-\omega_r+\chi^{-1}_{S}(\omega_{r},t)}.
\end{equation}
To apply the CFF method to the subsystem, it is essential to outline the experimental protocol for detecting $\chi_{S}(\omega_{r},t)$. As illustrated in Fig.~\ref{S1}, the subsystem $S$ (represented by the black circle) is coupled to subsystem $S^{\prime}$. By applying a real frequency ($\omega_{r}$) drive excited at site $i$, the component of the response function, i.e., $\chi_{S,ji}(\omega_{r},t)$, can be experimentally detected. Repeating this procedure while traversing all lattice sites within $S$ yields the complete information of $\chi_{S}(\omega_{r},t)$.
In this case, it is evident that $\lim\limits_{t\rightarrow \infty}\chi^{-1}_{S}(\omega_{r},t)=\omega_{r}-H_{S}+i\gamma-\Sigma_{S}(\omega_{r}+i\gamma)$, which leads to
\begin{equation}
\begin{split}
 \lim\limits_{t\rightarrow \infty}\mathscr{G}_{S,\omega_r}(t)&=G^{\mathrm{CFF}}_{S,\mathrm{eff}}(\omega_{c},\omega_{r})\\
&=\frac{1}{\omega_{c}+i\gamma-H_S-\Sigma_{S}(\omega_r+i\gamma)}, 
\end{split}
\end{equation}
where the double frequency Green's function defined above under OBC depends on the specific real frequency $\omega_r$, with the result under PBC given in Eq. 20. As discussed aforementioned, this result being an experimental observable is consistent with the Green's function under non-Hermitian approximation at $\omega_r$, exhibiting the non-Bloch response and showing distinct outcomes from CFE and CFS methods.

\section{Derivation of the subsystem's Green's function}
This section outlines the derivation of the analytic form of the subsystem's Green's function with a frequency-dependent self-energy. The discussion begins by considering the entire system under PBC along the $x$-direction and OBC along the $y$-direction. The analysis for OBC along the $x$-direction is straightforward by extension.

Notably, in the following derivation, we employ a Hermitian Fermionic system to derive the analytic form of the subsystem's Green's function. However, it is only the mathematical result (matrix representation of the Green's function) that is essential for investigating the response function solved from the bosonic-driven dissipative equation (Eq. 11 in the main text).

As introduced in the main text, we choose the one-dimensional Rice-Mele model as the subsystem Hamiltonian:
\begin{equation}
\hat{H}_{S}=\sum\limits_{k}\hat{\psi}^{\dag}_{k}H_{S}(k)\hat{\psi}_{k},
\end{equation}
where $\hat{\psi}^{\dag}_{k}=(\hat{c}^{\dag}_{kA},\hat{c}^{\dag}_{kB})$, $\hat{\psi}_{k}=(\hat{c}_{kA},\hat{c}_{kB})^{T}$ and $H_{S}(k)=(t_{1}+t_{2}\mathrm{cos}k)\sigma_{x}+t_{2}\mathrm{sin}k\sigma_{y}+(\lambda \mathrm{sin}k-\mu_{S})\sigma_{z}$.
For the subsystem $S^{\prime}$, with Fermion annihilation (creation) operators $\hat{d}_{\boldsymbol{R}}$ $(\hat{d}^{\dag}_{\boldsymbol{R}})$ at site $\boldsymbol{R}=(x,y)$, a Fourier transformation along the $x$-direction yields operators satisfying PBC along $x$-direction and OBC along $y$, for example: 
\begin{equation}
\hat{d}^{\dag}_{ky,A}=\frac{1}{\sqrt{N_{S}}}\sum\limits^{2N_{S}-1}_{x=\mathrm{odd}}e^{i k(x+1)/2}\hat{d}^{\dag}_{\boldsymbol{R}},\quad \hat{d}_{ky,B}=\frac{1}{\sqrt{N_{S}}}\sum\limits^{2N_{S}}_{x=\mathrm{even}}e^{-i kx/2}\hat{d}_{\boldsymbol{R}},
\end{equation}
where $x$ and $y$ are labeled as shown in Fig. 2 (a). Notably, although the subsystem $S^{\prime}$ is a single-band model, it is divided into parts $A$ and $B$ to align with subsystem $S$ for simplicity. The Hamiltonian for subsystem $S^{\prime}$ then becomes
\begin{equation}
\begin{split}
\hat{H}_{S^{\prime}}&=\sum\limits_{k}\sum\limits^{N_{y}+1}_{y=2}\hat{\Psi}^{\dag}_{ky}h_{S^{\prime}}(k)\hat{\Psi}_{ky}+\sum\limits_{k}\sum\limits^{N_{y}}_{y=2}(t_y\hat{\Psi}^{\dag}_{ky}\hat{\Psi}_{ky+1}+h.c.),
\end{split}
\end{equation}
where $\hat{\Psi}^{\dag}_{ky}=(\hat{d}^{\dag}_{ky,A},\hat{d}^{\dag}_{ky,B})$, $\hat{\Psi}_{ky}=(\hat{d}_{ky,A},\hat{d}_{ky,B})^{T}$ and $h_{S^{\prime}}(k)=(t_{x}+t_{x}\mathrm{cos}~k)\sigma_{x}+t_{x}\mathrm{sin}~k \sigma_{y}-\mu_{S^{\prime}}\sigma_0$. The $S$-$S^{\prime}$ coupling is 
\begin{equation}
\begin{split}
\hat{H}_{SS^{\prime}}&=\sum\limits_{x y }(T_{\boldsymbol{R}}\hat{c}^{\dag}_{x}\hat{d}_{\boldsymbol{R}}+h.c.),\\
\end{split}
\end{equation}
where
\begin{equation}
\begin{split}
\hat{c}^{\dag}_{x=\mathrm{odd}}&=\frac{1}{\sqrt{N_{S}}}\sum\limits_{k}e^{-ik (x+1)/2}\hat{c}^{\dag}_{kA},\quad
\hat{c}^{\dag}_{x=\mathrm{even}}=\frac{1}{\sqrt{N_{S}}}\sum\limits_{k}e^{-ik x/2}\hat{c}^{\dag}_{kB}, \\
\end{split}
\end{equation}
and $T_{x=\mathrm{odd},y=2}=t_{A}$, $T_{x=\mathrm{odd},y\geq2}=T_{x=\mathrm{even},y}=0$. 
Subsequently, we can define the subsystem's Green's function as
\begin{equation}
\begin{split}
G^{R}_{S,\alpha\beta}(k,t)&=-i\theta(t)\langle \{\hat{c}_{k\alpha}(t),\hat{c}^{\dag}_{k\beta}\}\rangle_{T}\\
G^{R}_{S^{\prime}S,\alpha\beta}(k,y,t)&=-i\theta(t)\langle \{\hat{d}_{ky,\alpha}(t),\hat{c}^{\dag}_{k\beta}\}\rangle_{T},\\
\end{split}
\end{equation} 
where $\langle ...\rangle_{T}$ represents thermal average. 
Here $\hat{O}(t)=e^{-i\hat{H}_{tot} t}\hat{O}e^{i \hat{H}_{tot}t}$ denotes the operator in the Heisenberg representation.

Using the commutation relations
\begin{equation}
\begin{split}
[\hat{c}_{k\alpha},\hat{H}_{tot}]&=\sum\limits_{\beta}[H_{S}(k)]_{\alpha\beta}\hat{c}_{k\beta}+\delta_{\alpha A}t_{A}\hat{d}_{k2,A},\\
[\hat{d}_{k2,\alpha},\hat{H}_{tot}]&=\sum\limits_{\beta}[h_{S^{\prime}}(k)]_{\alpha\beta}\hat{d}_{k2,\beta}+t_{y}\hat{d}_{k3,\alpha}+\delta_{\alpha A}t_{A}\hat{c}_{kA},\\
[\hat{d}_{ky,\alpha},\hat{H}_{tot}]&=\sum\limits_{\beta}[h_{S^{\prime}}(k)]_{\alpha\beta}\hat{d}_{ky,\beta}+(t_{y}\hat{d}_{ky-1,\alpha}+t_{y}\hat{d}_{ky+1,\alpha}),\quad y\geq 3
\end{split}
\end{equation}
the equations of motion for the Green's function in the frequency domain are:
\begin{equation}
\begin{split}
\omega G^{R}_{S}(k,\omega)&=\boldsymbol{I}+H_{S}(k)G^{R}_{S}(k,\omega)+\mathrm{Diag}(t_{A},0) G^{R}_{S^{\prime}S}(k,2,\omega),\\
\omega G^{R}_{S^{\prime}S}(k,2,\omega)&=h_{S^{\prime}}(k)G^{R}_{S^{\prime}S}(k,2,\omega)+T^{\dag}G^{R}_{S^{\prime}S}(k,3,\omega)+\mathrm{Diag}(t_{A},0)G^{R}_{S}(k,\omega),\\
\omega G^{R}_{S^{\prime}S}(k,n,\omega)&=h_{S^{\prime}}(k)G^{R}_{S^{\prime}S}(k,n,\omega)+TG^{R}_{S^{\prime}S}(k,n-1,\omega)+T^{\dag}G^{R}_{S^{\prime}S}(k,n+1,\omega),\quad (n\geq 3).\\
\end{split}
\end{equation} 
Here $T=t_{y}\boldsymbol{I}$ denote the inter-layer coupling in $S^{\prime}$.
Consequently, we obtain
\begin{equation}
G^{R}_{S}(k,\omega)=\frac{1}{\omega-H_{S}(k)+i\gamma-\Sigma_{S}(k,\omega+i\gamma)},
\end{equation}
where $\gamma>0$ is introduced in the standard form of the retarded Green's function. By defining a $H_{\mathrm{eff}}(k,\omega)$, the Green's function can be relabeled as $G^{R}_{S,\mathrm{eff}}(k,\omega)$, as given in Eq. 5 and Eq. 6 in the main text. Here, the self-energy is expressed as
\begin{equation}
\Sigma_{S}(k,\omega)=\mathrm{Diag}(\big[\frac{t^{2}_{A}}{\omega-h_{S^{\prime}}(k)-T\frac{1}{\omega-h_{S^{\prime}}(k)-Q_{N_{y}}(k,\omega)}T^{\dag}}\big]_{AA},0),
\end{equation}
with $Q_{2}(k,\omega)=0,Q_{n+1}(k,\omega)=T\frac{1}{\omega-h_{S^{\prime}}(k)-Q_{n}(k,\omega)}T^{\dag}$ for $n\geq 2$. This result is equivalent to $\Sigma^{AA}_{S}(k,\omega)=[\frac{t^{2}_{A}}{\omega-H_{S^{\prime}}(k)}]_{11}$ as expressed in the main text. Numerically, $Q_{N_{y}}(k,\omega)$ is obtained iteratively.

Furthermore, it is easy to check that the Green's function under OBC is 
\begin{equation}
G^{R}_{S,\mathrm{eff}}(\omega)=\frac{1}{\omega-H_{S}+i\gamma-\Sigma_{S}(\omega+i\gamma)},
\end{equation}
where the self-energy is given by 
\begin{equation}
\Sigma_{S}(\omega)=\frac{t^{2}_{A}}{\omega-h_{S^{\prime}}-T\frac{1}{\omega-h_{S^{\prime}}-Q_{N_{y}}(\omega)}T^{\dag}},
\end{equation}
where $h_{S^{\prime}}$ represents the real space matrix form of the one-dimensional chain along the $x$-direction in $S^{\prime}$, $Q_{2}(\omega)=0$, and $Q_{n+1}(\omega)=T\frac{1}{\omega-h_{S^{\prime}}-Q_{n}(\omega)}T^{\dag}$ for $n\geq 2$.

This comprehensive derivation provides a clear pathway to understanding the mathematical structure of the subsystem's Green's function, facilitating further analysis and interpretation.

\section{Poles of the complex frequency Green's function}
This section analyzes the pole structures of the complex frequency Green's function, comparing cases with and without non-Hermitian approximation.
For the exact case, the conventional poles for $G^{R}_{S,\mathrm{eff}}(k,\omega)$ are defined through the determinant equation:
\begin{equation}
\mathrm{det}[\omega-H_{S,\mathrm{eff}}(k,\omega)]=0.\label{poles}
\end{equation}
Denoting $\omega_{p}$ as solutions to Eq.~\ref{poles}, we establish that 
\begin{equation}
\omega_{p}(k)=\omega_{r}(k)-i\gamma,
\end{equation}
where $\omega_{r}\in \mathbb{R}$ satisfies
\begin{equation}
\mathrm{det}[\omega_{r}-H_{S}(k)-\Sigma_{S}(k,\omega_{r})]=0.
\end{equation}
This relationship emerges naturally because all poles of $G^{R}_{S,\mathrm{eff}}(k,\omega)$ belong to those of $G^{R}(k,\omega)$, which exhibit imaginary parts of $-i\gamma$. This conclusion naturally extends to the OBC case.

Subsequently, we discuss the eigenvalue of $H_{S,\mathrm{eff}}(k,\omega_{r})$ for $\omega_{r}\in\mathbb{R}$. The central conclusion is that if we denote $\varepsilon_{n}(k,\omega_{r})$ as the eigenvalue of $H_{S,\mathrm{eff}}(k,\omega_{r})$, then the inequality $\mathrm{Im}~\varepsilon_{n}(k,\omega_{r})<-\gamma$ holds. 
To derive this, consider the matrix decomposition $\frac{1}{X+iY}$ with Hermitian $X$ and positive-definite Hermitian $Y$. It is easy to check that 
\begin{equation}
(X+iY)(Y^{-1}X\frac{1}{XY^{-1}X+Y}-i \frac{1}{XY^{-1}X+Y})=I,
\end{equation}
where $XY^{-1}X+Y$ is Hermitian positie-definite and invertible. This leads to the transformation 
\begin{equation}
T\frac{1}{X+iY}T^{\dag}=X^{\prime}-iY^{\prime},
\end{equation}
where $T$ here represents a Hermitian matrix ($T=t_{y}\boldsymbol{I}$ in our case), $X^{\prime}=TY^{-1}X\frac{1}{XY^{-1}X+Y}T^{\dag}$ is Hermitian and $Y^{\prime}=T\frac{1}{XY^{-1}X+Y}T^{\dag}$ is Hermitian positive-definite.
Applying this framework to self-energy terms $Q_{m}(k,\omega_{r}+i\gamma)$ where $-\mathrm{Im}~Q_{m}(k,\omega_{r}+i\gamma)$ is Hermitian positive-definite, we identify $\omega_{r}-h_{S^{\prime}}(k)-\mathrm{Re}~Q_{m}(k,\omega_{r}+i\gamma)=X$ and $i\gamma-\mathrm{Im}~Q_{m}(k,\omega_{r}+i\gamma)=iY$, yielding
\begin{equation}
T\frac{1}{\omega_{r}+i\gamma-h_{S^{\prime}}(k)-Q_{m}(k,\omega_{r}+i\gamma)}T^{\dag}=T\frac{1}{X+iY}T^{\dag}=X^{\prime}-iY^{\prime}=Q_{m+1}(k,\omega_{r}+i\gamma),
\end{equation}  
where $-\mathrm{Im}~Q_{m+1}(k,\omega_{r}+i\gamma)$ remains Hermitian positive definite.
Since $-\mathrm{Im}~Q_{3}(k,\omega_{r}+i\gamma)=-\mathrm{Im}~\frac{t^{2}_{y}}{\omega_{r}-h_{S^{\prime}}(k)+i\gamma}$ is Hermitian positive definite, the iterative analysis confirms that $-\mathrm{Im}~\Sigma_{S}(k,\omega_{r}+i\gamma)$ maintains positive-definiteness in our model, enabling the decomposition 
\begin{equation}
H_{S,\mathrm{eff}}(k,\omega_{r})=H_{S}(k)+X^{\prime}-i(\gamma \boldsymbol{I} +Y^{\prime}).
\end{equation}
For normalized eigenstates $H_{S,\mathrm{eff}}(k,\omega_{r})\psi=\varepsilon(k,\omega_{r}) \psi$ with $\psi^{\dag}\psi=1$, we obtain
\begin{equation}
\begin{split}
\mathrm{Im}~ \psi^{\dag}H_{S,\mathrm{eff}}(k,\omega_{r})\psi&=\mathrm{Im}~\varepsilon(k,\omega_{r})\\
&=-\psi^{\dag}Y^{\prime}\psi-\gamma.
\end{split}
\end{equation}
Since $Y^{\prime}$ is positive definite, we conclude that $\mathrm{Im}~\varepsilon(k,\omega_{r})<-\gamma$. As the above derivation is irrelevant to the boundary condition, the conclusion holds under OBC, explaining the imaginary gap in the non-Hermitian approximation (light yellow region as shown in Fig. 4(a)).

\bibliography{supple-ref}
\bibliographystyle{apsrev4-1}